\documentclass[fleqn,amssymb,floats]{revtex4}

\usepackage{amsmath}
\usepackage{psfrag}
\usepackage{graphicx}
\usepackage{floatflt}

\def\Z{{\mathbb Z}}
\def\R{{\mathbb R}}
\def\la{\langle}
\def\ra{\rangle}

\newcommand{\be}{\begin{equation}}
\newcommand{\ee}{\end{equation}}
\newcommand{\bea}{\begin{eqnarray}}
\newcommand{\eea}{\end{eqnarray}}

\begin{document}


\title{Logarithmic scaling for height variables in the Abelian sandpile model}
\author{Geoffroy Piroux and Philippe Ruelle\footnote{Chercheur qualifi\'e FNRS}}
\affiliation{Universit\'e catholique de Louvain\\
Institut de Physique Th\'eorique\\
B--1348 \hskip 0.3truecm Louvain-la-Neuve, Belgium}
\date{\today}
\begin{abstract}
We report on the exact computation of the scaling form of the 1--point function, on the
upper--half plane, of the height 2 variable in the two--dimensional Abelian sandpile model. 
By comparing the open versus the closed boundary condition, we find that the scaling field
associated to the height 2 is a logarithmic scalar field of scaling dimension 2, belonging
to a $c=-2$ logarithmic conformal field theory. This identification is confirmed by numerical
simulations and extended to the height 3 and 4 variables, which exhibit the same scaling
form. Using the conformal setting, we make precise proposals for the bulk 2--point
functions of all height variables.
\end{abstract}

\maketitle
\section{Introduction}

Sandpile models are open dynamical systems which generically show a wide variety of
critical behaviours \cite{ip,dhar1}. In two dimensions, one may be tempted to analyse the
stationary state of a sandpile in terms of a local conformal field theory. As most
sandpile models have a number of unusual features, the underlying conformal theories are
expected to be peculiar as well. It is now established that at least some of them fall in
the class of logarithmic conformal theories (LCFT) \cite{flohr,gab}.

In the model we consider here, which is the model originally introduced by Bak, Tang
and Wiesenfeld \cite{btw}, also called the (isotropic) Abelian sandpile model \cite{dhar2},
a number of non--trivial features specific to a $c=-2$ LCFT have been successfully compared
with exact results \cite{md1,md2,bip,iv,mr,r,jeng1,jeng2,jeng3,pr1,pr2}, confirming the
relevance of the conformal description.

The Abelian sandpile model is most naturally formulated in terms of height variables.
These random variables, sitting at the sites of a lattice, take the values 1, 2, 3 and 4,
and evolve under a discrete time dynamics. When the lattice is finite, there is a unique
stationary measure, which is uniform on its support, formed by the so--called recurrent
configurations \cite{dhar2}. If the continuum limit of this measure is a $c=-2$
LCFT measure, each of the four height variables should converge to a definite conformal
field in the scaling limit, in such a way that the multisite probabilities are reproduced
by conformal correlation functions.

Early on, fundamental differences among the four height variables emerged. The
Bombay trick \cite{md1} allowed Majumdar and Dhar to handle the height 1 variable rather
easily, and led to the clear identification of its scaling form with a primary field
$\phi(z,\bar z)$ of dimensions (1,1) \cite{mr}. Technically the other three height
variables are harder. To date, the only known result, due to Priezzhev, is an exact
integral formula for the 1--site probabilities $P_{2}, P_{3}, P_{4}$ on the discrete plane
$\Z^{2}$ (or infinitely far from boundaries) \cite{priez}. As these numbers must be
subtracted from the random variables before making contact with the conformal fields (which
have a zero vacuum expectation value on the plane), they give no information on the fields
themselves. Even though different arguments have suggested a genuine difference 
between the height 1 and the heights 2, 3 and 4 in the bulk, the exact scaling form of the
latters remains an open question.

In this Letter, we provide an answer to this problem. The results which follow
show that the height 2 scales like a logarithmic conformal field $\psi(z,\bar z)$ of
dimensions (1,1) ---the logarithmic partner of the primary field $\phi$---, and relying on
numerical simulations of the lattice model, we  argue that this is also the scaling form for
the heights 3 and 4. Here we content ourselves with giving the main results of the lattice
calculations and simulations, and their comparison with the LCFT predictions. The details
of the analysis will be presented elsewhere.


\section{The height 2 on the upper--half plane}

Information about the scaling form of the height 2 variable requires the knowledge of
some of its correlations. The 1--site probability $P_{2}(z)$ on the upper--half plane is a
disguised correlator of the height 2 with itself (its mirror image), and gives the simplest
way to assess its scaling behaviour. It can be computed in essentially the same way the 1--site
probability $P_{2}=\lim_{z \to i\infty}P_{2}(z)$ has been computed on the full plane by
Priezzhev \cite{priez}.

The height variables are the most basic microscopic variables, but they are
strongly correlated over the whole lattice because they must make up recurrent
configurations. A useful alternative is to treat the recurrent configurations as a global
random variable, uniformly distributed, and in turn, the recurrent configurations 
can be traded for the set of spanning trees on the lattice one considers \cite{md2}. While
the spanning tree picture may be avoided for calculations with heights 1, it seems to be
unavoidable for the height 2 (and 3 and 4). 

Being interested in height observables, one should translate the
height values at certain sites into well--defined characteristics of the spanning trees.
This can be done generally, but for what follows, we only discuss the case of a single
height 2. In a spanning tree, any site belonging to the branch(es) grown from a site $i$ is
called a predecessor of $i$. It can then be shown \cite{priez} that a given recurrent
configuration has a height 2 at site $i$ if and only if, in the corresponding spanning
tree, the site $i$ has exactly one predecessor among its nearest neighbours. The
distribution on spanning trees being uniform, one has to count the spanning trees which
satisfy this (non--local) characterization at $i$.

Priezzhev has devised a method to compute this number \cite{priez}. It involves
counting the trees with a certain local property, and also graphs with some other
non--local properties. The former can be done by standard methods of graph theory, like
the introduction of a defect matrix, the calculation then boiling down to a small
determinant. The latter calculation requires to enumerate the graphs with a loop (a
non--local constraint but easily computed using a defect matrix) and also 
what Priezzhev called $\Theta$--graphs, which are graphs with an imposed $\Theta$--shaped
circuit. Both non--local contributions are divergent, but the combination which is relevant to
the site probabilities is finite.

We have used this technique to compute the probability to have a height 2 somewhere on the
upper--half plane. With respect to the plane, this situation brings a few changes.

The most obvious change is the presence of the boundary, on which boundary conditions
must be prescribed. Two natural --and the only ones known-- boundary
conditions may be imposed along the boundaries, open (Dirichlet) and closed (Neumann).
Their precise definition is not important for what follows and can be found for instance in
\cite{pr1}. More important is the fact that the operator which converts an open boundary
condition into a closed one, or vice--versa, has been identified as a $c=-2$ primary field
$\mu(z)$ of weights $-{1 \over 8}$ \cite{r}. It will be a crucial
ingredient in our proof that the height 2 variable scales like the logarithmic field
$\psi$. Indeed the probability that a site have height 2 will be computed for both an open
and a closed boundary condition, but the two are not independent since they can be
related to each other by the use of the field $\mu$. This is a non--trivial
consistency check that involves the 4--point function with two $\mu$'s and two $\psi$'s.

The counting of $\Theta$--graphs is also a little bit more complicated on the upper--half
plane than on the full plane. The rotational invariance is lost and the translational symmetry
is broken in one direction. Moreover the way the $\Theta$--circuit can touch the
boundary has to be controlled. These circumstances increase the number of different
graph configurations that have to be computed separately, and complicate the
expressions. 

The expressions we have obtained for the probabilities using the $\Theta$--graph
technique are rather cumbersome, to the extent that reproducing them here is unlikely to
be helpful. The local and the loop contributions are simple but those
coming from the $\Theta$--graphs are fairly long. They involve six--fold integrals and a
summation over the sites of the upper--half plane. The horizontal translational invariance
enables one to carry out the infinite sum over the horizontal coordinate, which in turn
allows to do one of the six integrations (on the full plane, the two infinite
summations can be done, and then two of the six integrations, leaving a four--fold
integral \cite{priez}). The integrand/summand involves the determinants of 4--by--4
matrices, and contain a rather complicated dependence in $m$, the distance to the
boundary of the reference site. 

The asymptotic analysis of these expressions is long and technical, but yields the
asymptotic form of $P^{\rm op}_{2}(m)$ and $P^{\rm cl}_{2}(m)$, the probability that a 
site at a distance $m$ from the boundary has a height equal to 2, when the boundary is
all open or all closed. We have obtained the following results,
\bea
P_{2}^{\rm op}(m) - P_{2} &=& 
{1\over m^2} \big(a+{b\over2}+b\log m\big)+\mathcal O(m^{-3}\log^{k}\!m),
\label{p2op}\\
P_{2}^{\rm cl}(m) - P_{2} &=& 
-{1\over m^2} \big(a+b\log m\big)+\mathcal O(m^{-3}\log^{k}\!m).
\label{p2cl}
\eea
The height 2 probability $P_{2}$ on the full plane is known exactly as a multiple
integral, whose numerical evaluation yields $P_{2} = 0.1739$ \cite{priez}. The constant $a$
is also given in terms of integrals which we could not to evaluate analytically, but a
numerical integration gives $a = 0.0403$. As to the other constant, we found the exact
value $b=  {\textstyle{1 \over 4}} P_{1} = {1\over 2\pi^2}(1-2/\pi)$, with $P_{1} =
0.07363$ the probability that a site in the plane has height 1 \cite{md1}.

These results give a clear signal that the scaling field describing the height 2 variable
has scale dimension 2 and has a logarithmic component. The natural candidate in the $c=-2$
LCFT is the field $\psi$, namely the logarithmic partner of the primary field $\phi$,
with conformal dimensions $(1,1)$. 

If $\psi$ is indeed the scaling field of the height 2, the above dominant terms for the
subtracted probabilities should be equal to $\la \psi(z,z^*) \ra_{\rm op}$ and $\la
\psi(z,z^*) \ra_{\rm cl}$ with $z - z^{*} = 2im$. The non--chiral 1--point functions
on the upper--half plane are also equal to the chiral correlators on the full
plane, of two chiral $\psi$ inserted at $z$ and $z^*$, provided the coefficients
defining the most general chiral 2--point function are properly chosen \cite{cardy}. 

In order to be able to relate the open and the closed boundary conditions, we consider
the more general expectation value of $\psi$ on the upper--half plane, for which the
boundary condition on the real axis is all open except on the interval $I=[z_1,z_2]$
where it is closed. It can be expressed as a chiral 4--point function on the plane as
\be
\la \psi(z,z^*)\ra_{\rm UHP}^{{\rm op},I} = N_{\mu}^{-1}
z_{12}^{-1/4} \; \la \mu^{D,N}(z_{1})\mu^{N,D}(z_{2})\psi(z)
\psi(z^*)\ra_{\rm chiral},
\label{4pt}
\ee 
where $N_{\mu} = 1.18894$ normalizes the lattice 2--point functions of the fields $\mu$ \cite{r}.
In the sandpile model, it yields the (scaling limit of the) probability that
the site $z$ has a height equal to 2 in presence of a boundary which is open on ${\R}
\setminus I$ and closed on $I$. The two lattice results (\ref{p2op}) and (\ref{p2cl})
are then easily recovered upon taking $z_1,z_2 \to 0$ and $-z_1,z_2 \to +\infty$
respectively. In these limits, one may invoke the operator product expansion of two
$\mu$'s, which closes on different fields in the two cases, since the remaining boundary
condition is different
\cite{pr1},
\be
\mu^{D,N}(z_{1}) \mu^{N,D}(z_{2}) \longrightarrow  
\begin{cases}
N_{\mu} z_{12}^{1/4}\,{\mathbb I}_{D} + \ldots & \hbox{when $z_{1},z_{2} \to 0$},\\
\noalign{\smallskip}
N_{\mu} z_{12}^{1/4}\,[\omega_{N}(\infty) - {1 \over \pi} {\mathbb I}_{N} 
\log{z_{12}} + \ldots] & \hbox{when $-z_{1},z_{2} \to +\infty$,}
\end{cases}
\label{ope}
\ee
where ${\mathbb I}_{D}$ and ${\mathbb I}_{N}$ are the identity fields on an open resp.
closed boundary and $\omega_{N}$ is the logarithmic partner of ${\mathbb I}_{N}$. Thus
we obtain the two expectation values on the upper--half plane as
\bea
\la \psi(z,z^*) \ra_{\rm op} &=& \la \psi(z) \psi(z^*) \ra_{\rm chiral}, \\
\la \psi(z,z^*) \ra_{\rm cl} &=& \la \psi(z) \psi(z^*) \omega_{N}(\infty)
\ra_{\rm chiral} - {1 \over \pi} \lim_{-z_{1},z_{2} \to +\infty} \log{z_{12}} \, 
\la \psi(z) \psi(z^*) \ra_{\rm chiral}.
\label{1pt}
\eea

Because the primary field $\mu$ is degenerate at level 2, the chiral correlator $\la
\mu(z_{1}) \mu(z_{2}) \psi(z_{3}) \psi(z_{4}) \ra$ satisfies a second--order
differential equation, but the logarithmic nature of $\psi$ complicates the matter. It
is the logarithmic partner of a primary field $\phi$ of dimensions (1,1), into which it
transforms under a dilatation, and it is not quasi--primary, because it transforms into
primary fields $\rho(z,\bar z)$ and $\bar \rho(z,\bar z)$ of dimensions (0,1) and (1,0)
under special conformal transformations \cite{gk}. The operator product expansion
of $\psi$ with the chiral stress--energy tensor reads
\be
T(z)\psi(w,\bar w) = {\rho(w,\bar w) \over (z-w)^3} + 
{\psi(w,\bar w) - {1 \over 2} \phi(w,\bar w) \over (z-w)^2} + 
{\partial \psi(w,\bar w)\over z-w} + \ldots.
\label{T}
\ee
As a consequence, the differential equation satisfied by $\la \mu(z_{1}) 
\mu(z_{2}) \psi(z_{3}) \psi(z_{4}) \ra$ is inhomogeneous \cite{flohr}, where the 
inhomogeneity depends on the four other 4--point functions $\la \mu \mu \rho \psi \ra,
\, \la \mu \mu \phi \psi \ra,\, \la \mu \mu \psi \rho \ra$ and $\la \mu \mu \psi 
\phi \ra$. These 4--point correlators themselves satisfy inhomogeneous 
differential equations, and depend on yet four other correlators, $\la \mu \mu \rho \rho
\ra, \, \la \mu \mu \phi \rho \ra,\, \la \mu \mu \rho \phi \ra$ and $\la \mu \mu \phi 
\phi \ra$, which involve primary fields only, and hence may be computed independently, as
solutions of homogeneous differential equations. 

Altogether, there is a sequence of nine correlators to compute, the last one being $\la
\mu \mu \psi \psi \ra$. The five correlators which contain a $\phi$ satisfy a first
order differential equation, because the $\phi$ is also degenerate at level 2, so each
depends on one arbitrary coefficient. The other four satisfy second order equations, and
depend on two arbitrary coefficients. Thus the general 4--point function $\la
\mu(z_{1}) \mu(z_{2}) \psi(z_{3}) \psi(z_{4}) \ra$ depends on 13 arbitrary coefficients.

To fix their values, we note, from the operator product expansion (\ref{ope})
in the limit $z_{1},z_{2} \to 0$, that none of the nine correlators mentioned above can
have a logarithmic singularity, and that the resulting function $\la \psi(z,z^{*}) \ra_{\rm
op}$ must be a function of $z-z^*$ only. These two facts together fix 10 coefficients. One
more coefficient is fixed if one imposes that the limit $-z_{1},z_{2} \to +\infty$
remains regular, {\it i.e.} the logarithmic singularity is absent. As is clear from
(\ref{1pt}), this implies that $\la \psi(z,z^*) \ra_{\rm cl}$ is related, through the
image method, to a chiral field $\psi$ satisfying $\la \psi(z) \psi(z^{*}) \ra = 0$.
The same is true for the height 1 variable and its field $\phi$ \cite{pr1}.

All this leaves just two free parameters, themselves determined by demanding that the
limit $z_{1},z_{2} \to 0$ of (\ref{4pt}) reproduces the lattice result (\ref{p2op}). It
leads to our final result for the expectation value of $\psi$ on the UHP with a boundary
closed on $[z_{1},z_{2}]$ and otherwise open, 
\be
\la \psi(z,z^{*})\ra_{\rm UHP}^{{\rm op},I}
= {1\over (z-z^*)^2} {x-2\over\sqrt{1-x}} \Big\{ 2b\log\Big|{z-z^* \over 2}\Big| 
+ 2a + {b \over 2} - {b \over 4} {x-2 \over \sqrt{1-x}} 
+ {b (z-z^*)^{2} \over  (z_{1}-z^*)(z_{2}-z)}
\Big[ {1 \over x-2} + {1 \over 2\sqrt{1-x}} \Big] \Big\},
\label{p2I}
\ee
where $x = {z_{12}z_{34} \over z_{13}z_{24}}$ is a cross--ratio of the four insertion
points ($z_{3}=z_{4}^*=z$). One readily checks that in the limit $z_{1},z_{2} \to 0$, $x
\to 0$ and $\sqrt{1-x} \to +1$, the previous formula reduces to (\ref{p2op}).

To make the connection with the probability of a height 2 in presence of a closed
boundary, one takes $-z_{1},z_{2} \to +\infty$. Again $x$ goes to 0, but
$\sqrt{1-x}$ now tends to $-1$ \cite{pr1}. Indeed if we set $z_{1} = -R$ and $z_{2} = R$,
we find that $1-x = {(R-z)(R+z^{*}) \over (R-z^{*})(R+z)}$ has unit modulus and draws
a complete circle around 0 when $R$ varies from 0 to $+\infty$. This non--trivial
monodromy factor brings a global change of sign in (\ref{p2I}) with respect to the previous
case ($z_{1},z_{2} \to 0$), and the non--logarithmic $b$ terms drop out inside the curly
brackets (the last term does not contribute in either case). The resulting 1--point
function coincides exactly with the lattice result (\ref{p2cl}), thereby confirming in a
non--trivial way the identification of $\psi$ as the scaling field for the bulk height 2
variable in the sandpile model. We will add further support in the next section.

We finish this section by giving the expectation value of $\psi$ in the upper--half
plane, in presence of a boundary closed on $\R_{-}$ and open on $\R_{+}$, a formula we
will use in the next section. In this case, $z_{1} \to -\infty$, $z_{2} \to 0$, and a
simple calculation from (\ref{p2I}) yields
\be
\la \psi(z,z^{*}) \ra_{\rm cl,op} = -{z+z^{*} \over |z| (z-z^{*})^2} 
\Big\{ 2b\log\Big|{z-z^* \over 2}\Big|  + 2a + {b \over 2} + {b \over 4} 
{z+z^{*} \over |z|}\Big\}.
\label{clop}
\ee


\section{Simulations on a strip}

We have so far verified the identification of the height 2 variable with the field $\psi$
in the two extreme cases, namely in the upper--half plane with a homogeneous boundary
condition on the real line, either all open or all closed. It may be tested more
deeply by seeing how the profile of the height 2 probability varies in a region where
the boundary condition changes. The formula (\ref{clop}) corresponds to such a
situation, with half the real line open and the other half closed, and would describe the
way the probability $P_{2}(z)$ is interpolated between its open boundary value and its
closed boundary value.

We could however not carry out the lattice analysis for this case as we did for the
homogeneous boundary condition, so that the exact lattice result is not available.
Thus we turned to numerical simulations of the model to measure the lattice profile of
$P_{2}(z)$, with the extra advantage that the two other probabilities $P_{3}(z)$ and
$P_{4}(z)$ can be obtained with no more effort. 

Clearly the simulations can only be done on a finite array, in our case, a rectangle of
base $M$ and height $N$. We have imposed an open boundary condition on the bottom
boundary, a closed boundary condition on the top boundary, and we have evaluated the
four probabilities $P_{1}(n),\,P_{2}(n),\,P_{3}(n)$ and $P_{4}(n)$ on the vertical line 
joining the bottom to the top boundary, and located at mid--length (we have included 
$P_{1}(n)$ for completeness and for comparison purposes). For these results to be
significant (with respect to the CFT approach), both the height $N$ and the ratio $\tau=M/N$
must be large enough so that the scaling regime is approached, and so that the rectangle
can be considered as a good approximation of the infinite strip. We found that the values
$N=50$ and $\tau=4$ were sufficient, by verifying that larger values did not bring any
noticeable changes in the curves. To speed up the calculations, we chose open boundary
conditions on the left and right boundaries.

We have determined the 1--site probabilities $P_{i}(n)$ on a sample of nearly $17 \cdot
10^9$ recurrent configurations, generated from 1680 random recurrent configurations in the
following way. Applied to each of these, the dynamics of the model
produces a new recurrent configuration at each time step.  The first $10^{5}$ ones are 
not sampled, and from then on, every 50th configuration is sampled, for a total of
$10^{7}$ sampled configurations for each initial configuration. At each point $n$ of
the line joining the two boundaries, the probabilities are the ratios of the number of
occurrences by the size of the sample. From these numbers, we should subtract in each case
the probability $P_{k}$ in the bulk (far from the boundaries) before making the comparison
with the field--theoretic results. The most distant sites from the boundaries are the
central sites, but because the strip is finite, the lattice probabilities at the central
sites have not yet attained their limit value $P_{k}$. So we have subtracted an
effective probability $P_{k}^{\rm eff}$, computed by a fitting procedure. The
results are shown as colour dots on the left plots of Figures 1 and 2. 

Despite the fact that the size of the sample ($\sim 10^{10}$) is desperately small
compared to the volume of the phase space ($\sim 10^{5\,045}$ !), the
fluctuations in the data are surprisingly small, and follow a normal distribution to a
good approximation. Based on an analysis of these, we expect errors less than
$2\%$ for $P_{1}(n)$, less than $1\%$ for $P_{2}(n)$ and $P_{3}(n)$, and less than
$0.5\%$ for $P_{4}(n)$. The precision naturally increases for more frequent events
(the bulk values are $P_{1} = 0.0736, P_{2} = 0.1739, P_{3} = 0.3063, P_{4}
= 0.4461$ \cite{priez}).

\begin{figure}[hbt]
\psfrag{n}{$n$}
\psfrag{Pa}{}
\psfrag{Loga}{}
\includegraphics[scale=1.05]{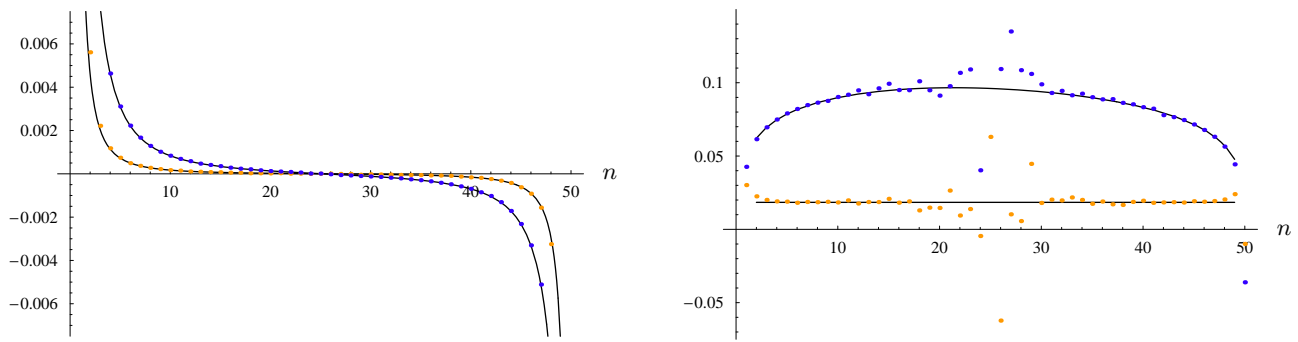}
\caption{Subtracted probabilities to find a height 1 (in yellow) or a height 2 (in blue)
at a site lying along the short medial line of a rectangle $M \times N$, and going from
an open boundary (left end of the graph) to a closed boundary (right end). The dots
correspond to data obtained from numerical simulations, with $M=200$ and $N=50$, while the
solid curves represent the conformal predictions for the limit $M=+\infty$. On the left are
plotted the raw data and the functions (\ref{p2cyl}) and (\ref{p1cyl}) to which they can be
directly compared. On the right, the same plots, but multiplied by the factor $\big({L\over
\pi}\big)^{\!2} {\sin^2(\pi n/L)\over \cos(\pi n/L)}$, reveal the presence or the
absence of a logarithmic dependence.}
\end{figure}

The results pertaining to the height 2 probability $P_{2}(n)$ may be directly compared with
the probability profile on the infinite strip obtained from the LCFT picture, on the basis
that the height 2 variable goes to the field $\psi$ in the scaling limit. For that,
the corresponding curve on the upper--half plane is simply transformed to the infinite
strip, of width $L$, by the conformal transformation $w = {L \over \pi} \log{z}$.

By integrating the infinitesimal transformation of $\psi$ given by (\ref{T}), one finds
its finite transformation under $z \to w$,
\be
\psi_{\rm new}(w,\bar w) = |z'(w)|^{2} \Big\{\psi_{\rm old}(w,\bar w) 
- {1 \over 2} \log |z'(w)|^2 \, \phi_{\rm old}(w,\bar w)
+ {z''(w) \over 2z'^{2}(w)}\, \rho_{\rm old}(w,\bar w)
+ {\bar z''(\bar w) \over 2\bar z'^{2}(\bar w)} \, 
\bar \rho_{\rm old}(w,\bar w)\Big\}.
\ee
Under the above mapping $z=\exp{(\pi w/L)}$ of the upper--half plane onto the infinite
strip, and taking expectation values, one obtains $\la \psi(w,\bar w) \ra_{\rm strip}$
in terms of $\la \psi(w,\bar w) \ra_{\rm UHP}$, $\la \phi(w,\bar w) \ra_{\rm UHP}$,
$\la \rho(w,\bar w) \ra_{\rm UHP}$ and $\la \bar \rho(w,\bar w) \ra_{\rm UHP}$. 
The expectation of $\psi$ is given in (\ref{clop}), to which that of $\phi$ is related
through a dilatation, whereas the expectations of $\rho$ and $\bar\rho$ can be shown to
vanish,
\be
\la \phi(z,z^{*}) \ra_{\rm cl,op} = -{2b (z + z^{*}) \over (z-z^{*})^{2} |z|}, 
\qquad \la \rho(z,z^{*}) \ra_{\rm cl,op} = \la \bar\rho(z,z^{*}) \ra_{\rm cl,op} = 0.
\label{phi}
\ee

Putting all together, we obtained the following formula for the expectation value of
$\psi$ on an infinite strip of width $L$, and coordinates $w = u+iv$,
\be
\la\psi(w,w^{*})\ra_{\rm strip}^{\rm op,cl}
= \Big({\pi\over L}\Big)^{\!2} {\cos(\pi v/L)\over\sin^2(\pi v/L)}
\Big\{ a + {b\over4} \Big[1+\cos\big({\pi v\over L}\big)\Big] 
+ b \log\!\Big({L\over\pi}\sin\big({\pi v\over L}\big)\Big)  \Big\}.
\label{p2cyl}
\ee
The open boundary is at $v=0$ and the closed one at $v=L$.

The height 1 variable scales exactly to the primary field $\phi$, normalized as in
(\ref{phi}) \cite{pr1} (that is why we introduced the factor $-{1 \over 2}$ in
(\ref{T})). So one quickly obtains
\be
\la\phi(w,w^{*})\ra_{\rm strip}^{\rm op,cl} = b
\Big({\pi\over L}\Big)^{\!2} {\cos(\pi v/L)\over\sin^2(\pi v/L)}.
\label{p1cyl}
\ee

The two functions of $v$ in (\ref{p2cyl}) and (\ref{p1cyl}) have been plotted in the
left graph of Figure 1, as solid lines. In both cases, we have chosen $L=N+{1 \over
2}$, for the following reason. In the continuum, $v=L$ is the position of the closed
boundary, on which the Neumann condition is enforced. On the lattice, this condition is
imposed through a mirror symmetry about the line $n=N+{1 \over 2}$, so that it is at that
point that the normal derivative vanishes. 

One sees that the conformal curves are very close to the profiles obtained
from numerical simulations, adding much support to the identification of $\psi$ as the
scaling field for the height 2 variable. Because $N = 50$ is relatively small, the plots with
$L=N$ would be slightly shifted near the closed side, resulting in a visible discrepancy at
the right end of the plots. The gross features of the two curves
are essentially the same and do not reveal any qualitative difference between the height 1
and the height 2, namely the logarithmic nature of the latter. In order to make the
difference to appear more clearly, we have plotted, in the right graph of Figure 1, the
same curves and the same data but divided by the damping prefactor $\big({\pi\over L}
\big)^{\!2} {\cos(\pi n/L)\over\sin^2(\pi n/L)}$, which tends to make the
function vanish precisely for the values of $v$ where the logarithm becomes important.
One then sees that the plot for the height 1 is flat, while the other one shows a non--trivial
dependence, and agrees well with the simulations (the discrepancy in the central region is
due to the difference between the true bulk value of $P_{1}$ and $P_{2}$ and the
effective values used in the subtraction, a difference which is amplified by the
factor $L^{2}/(\pi^{2}\cos{(\pi v/L)})$; the same factor also amplifies any disparity
between the dots and the curve). We note that for the height 2, the curve is
asymmetric around the middle point, a consequence of the extra factor $b \over 2$ in the 
probability near an open boundary, see (\ref{p2op}) and (\ref{p2cl}).

\begin{figure}[hbt]
\psfrag{n}{$n$}
\psfrag{Pb}{}
\psfrag{Logb}{}
\includegraphics[scale=1.05]{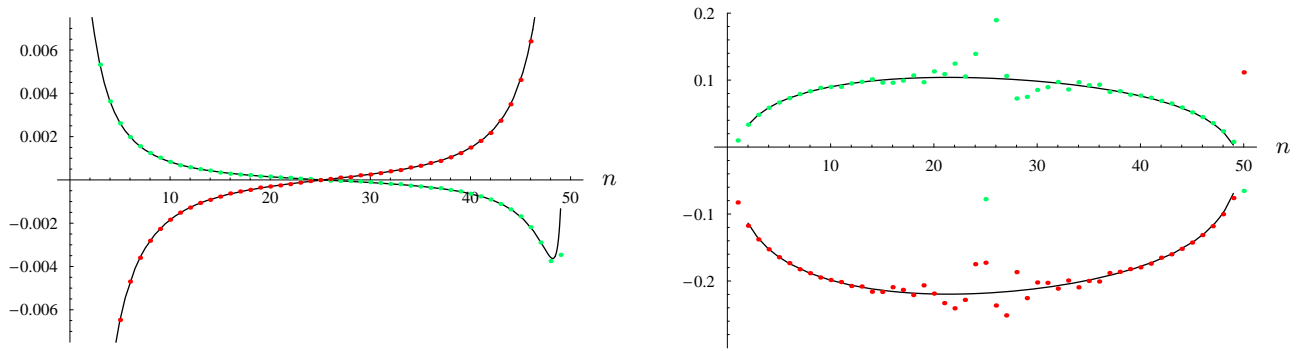}
\caption{Subtracted probabilities to find a height 3 (in green) or a height 4 (in red)
on a line joining the open boundary to the closed boundary of a strip, as in Figure 1.}
\end{figure}

The graphs in Figure 2 concern the height 3 and height 4 probabilities. As before
the dots in the left plot represent the subtracted probabilities obtained from numerical
simulations. Suspecting that the height 3 and 4 variables scale the same way as the height
2, we have fitted these data with the expressions they would follow if that were true,
namely the expression (\ref{p2cyl}). In other words, we have assumed that the probability
profiles for the heights 3 and 4 across the finite discrete strip of width $N=L-{1 \over
2}$ have the form
\be
P_{i}(n) - P_{i}^{\rm eff} = \Big({\pi\over L}\Big)^{\!2} {\cos(\pi n/L)\over
\sin^2(\pi n/L)} \Big\{ a_{i} + {b_{i}\over4} \Big[1+\cos\big({\pi n\over L}\big)
\Big]  + b_{i} \log\!\Big({L\over\pi}\sin\big({\pi n\over L}\big)\Big)  \Big\},
\qquad i=3,4. 
\label{p34cyl}
\ee

The fit of the three parameters, performed on the 46 most central values of $n$, yields
the following values in each case,
\bea
P_3^{\rm eff} &=& 0.3063\,, \quad a_3 = -0.01243\,, \quad b_3 = 0.03810\,,
\label{fit3}\\
P_4^{\rm eff} &=& 0.4461\,, \quad a_4 = -0.04636\,, \quad b_4 = -0.05667\,.
\label{fit4}
\eea
By comparison, a fit of the data for the height 2 yields $b_{2} = 0.01896$
for the parameter controlling the logarithmic dependence (the exact value is $b = {P_{1}
\over 4} = 0.01841$), and $b_{1} = -0.000388$ for the height 1 (the exact value is 0).

Using the fitted values of the parameters, the curves (\ref{p34cyl}) are plotted on the
left graph of Figure 2, as solid lines. The rather large values of $b_{3}$ and $b_{4}$,
as well as the good agreement between the solid curves and the data, are a very strong
indication that the height 3 and 4 variables possess the same logarithmic scaling as the
height 2 variable. The plots in the right part of Figure 2 bear the same relation to the
left plots as those in Figure 1, and exhibit a clear logarithmic signature.

If the height 3 and 4 variables scale like the height 2 variable, their associated fields
must be $\psi_{i} = \alpha_{i} \psi + \beta_{i} \phi$, for $i=3,4$. Using the above
formulae for the expectation values and the fitted values in (\ref{fit3}) and (\ref{fit4}),
we obtain the relations $b_{i} = b \alpha_{i}$ and $a_{i} = a \alpha_{i} + b
\beta_{i}$, from which we deduce the values $\alpha_{3} = 2.07,\, \beta_{3} = -5.21,\,
\alpha_{4} = -3.08,\, \beta_{4} = 4.22$. 

In summary, we propose the following field assignments for the scaling limit of the four
height variables
\bea
{\rm height\ 1} &:& \quad h_1(z,\bar z) = \phi(z,\bar z)\,, \qquad \quad
{\rm height\ 3} : \quad  h_3(z,\bar z) = \alpha_{3}\psi(z,\bar z) + 
\beta_{3}\phi(z,\bar z)\,,\\
{\rm height\ 2} &:& \quad h_2(z,\bar z) = \psi(z,\bar z)\,, 
\qquad \quad {\rm height\ 4} : \quad h_4(z,\bar z) = \alpha_{4}\psi(z,\bar z) +
\beta_{4}\phi(z,\bar z)\,.
\eea
The numerical values are consistent with the sum being zero.


\section{Height correlations in the bulk}

The results we obtained in the previous sections can finally be used to make definite
statements about the height correlations on the plane. As far as we know, and apart from
those involving heights 1 exclusively, these correlations have never been computed nor
measured from simulations. 

The correlation of two distant heights $h_{i},h_{j}$, located at $z_{1},z_{2}$ on the
plane, is given by the 3--point function of the corresponding fields with the field
$\omega$, the logarithmic partner of the identity, inserted at infinity,
\be
P_{ij}(z_{12}) - P_i P_j = \la h_i(z_{1},\bar z_{1}) 
h_j(z_{2},\bar z_{2}) \omega(\infty) \ra, \qquad |z_{12}| \gg 1, \qquad i,j=1,2,3,4.
\ee
For this one needs the 3--correlators $\la \phi \phi \omega \ra$, $\la \phi
\psi \omega \ra$ and $\la \psi \psi \omega \ra$, the last one anyway requiring the
other two. 

The most general form of the 3--point function $\la \psi \psi \omega \ra$ depends on 10
arbitrary constants. Placing the $\omega$ at infinity and requiring that the result be
translationally invariant reduces it to 
\be
\la \psi(1) \psi(2) \omega(\infty) \ra = {1 \over |z_{12}|^{4}} \Big\{
C + 2B \log|z_{12}| + A \log^{2}|z_{12}| \Big\},
\ee
but also enforces the following correlators 
\bea
&& \hspace{-5mm} \la \phi(1) \phi(2) \ra = \la \psi(1) \phi(2) \ra = \la \psi(1) 
\psi(2) \ra = 0\,,\\
&& \hspace{-5mm} \la \phi(1) \phi(2) \omega(\infty) \ra = {A \over |z_{12}|^{4}}\,, \qquad
\la \phi(1) \psi(2) \omega(\infty) \ra = {1 \over |z_{12}|^{4}}\Big\{B + A 
\log|z_{12}|\Big\}.
\eea

Interestingly, the constant $A$ governs the long distance behaviour of all 2--site
height correlations, and is known form the calculation of the 2--site probability for 
heights 1, $A = -P^{2}_{1}/2$ \cite{md1}. Granting the field assignments made above for
the heights 3 and 4, we conclude that the leading terms of the 2--site probabilities on
the plane are given by
\be
P_{11}(r) - P_{1}^{2} \simeq -{P^{2}_{1} \over 2r^{4}}, \quad P_{1i}(r) - P_{1}P_{i} 
\simeq -{\alpha_{i}P^{2}_{1} \over 2r^{4}}\log{r}, \quad
P_{ij}(r) - P_{i}P_{j} \simeq -{\alpha_{i}\alpha_{j}P^{2}_{1} \over 2r^{4}}
\log^{2}{r}, \qquad i,j=2,3,4,
\ee
with $\alpha_{2}=1$, $\alpha_{3}=2.07$ and $\alpha_{4}=-3.08$. All height variables
are anticorrelated, except the height 4 which is positively correlated to the other three.
These tendencies are precisely those computed for the four height variables on an open
boundary \cite{iv}. 


\end{document}